# Live Multi-language Development and Runtime Environments

Fabio Niephaus[a], Tim Felgentreff[a,b], Tobias Pape[a], Robert Hirschfeld[a], and Marcel Taeumel[a]

a   Hasso Plattner Institute, University of Potsdam, Germany
b   Oracle Labs Potsdam, Germany

**Abstract**   *Context:* Software development tools should work and behave consistently across different programming languages, so that developers do not have to familiarize themselves with new tooling for new languages. Also, being able to combine multiple programming languages in a program increases reusability, as developers do not have to recreate software frameworks and libraries in the language they develop in and can reuse existing software instead.
*Inquiry:* However, developers often have a broad choice of tools, some of which are designed for only one specific programming language. Various Integrated Development Environments have support for multiple languages, but are usually unable to provide a consistent programming experience due to different language-specific runtime features. With regard to language integrations, common mechanisms usually use abstraction layers, such as the operating system or a network connection, which are often boundaries for tools and hence negatively affect the programming experience.
*Approach:* In this paper, we present a novel approach for tool reuse that aims to improve the experience with regard to working with multiple high-level dynamic, object-oriented programming languages. As part of this, we build a multi-language virtual execution environment and reuse Smalltalk's live programming tools for other languages.
*Knowledge:* An important part of our approach is to retrofit and align runtime capabilities for different languages as it is a requirement for providing consistent tools. Furthermore, it provides convenient means to reuse and even mix software libraries and frameworks written in different languages without breaking tool support.
*Grounding:* The prototype system Squimera is an implementation of our approach and demonstrates that it is possible to reuse both development tools from a live programming system to improve the development experience as well as software artifacts from different languages to increase productivity.
*Importance:* In the domain of polyglot programming systems, most research has focused on the integration of different languages and corresponding performance optimizations. Our work, on the other hand, focuses on tooling and the overall programming experience.

## The Art, Science, and Engineering of Programming

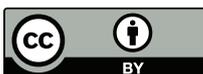

Live Multi-language Development and Runtime Environments

## 1 Introduction

High-level dynamic programming languages provide different abstractions to increase productivity, thus reducing software development costs. Although their dynamic properties sometimes come with performance costs, we observe that many developers prefer to use dynamic programming languages and are willing to trade performance – even when it matters – for a better programming experience.

Nonetheless, the fact that there are many different languages with different features has additional consequences for developers. In general, they often have to choose a language for their programs and are then limited to the features of that language and the libraries and frameworks developed in it. Additionally, they also have to learn how to use the tools for each language which can be a significant overhead.

In this paper, we propose an approach that aims to further improve the programming experience with regard to working with different high-level dynamic languages. To demonstrate this, we have implemented Squimera, an Integrated Development Environment (IDE) for dynamic languages. Squimera is named after Squeak/Smalltalk, an interactive programming system, and Chimera, a hybrid creature from Greek mythology.

**The Need for Software Development Tools**   Creating software is a comprehensive activity which requires knowledge, creativity, and practice, as well as a suitable set of software development tools. These tools are often provided as part of an IDE and usually support developers throughout the entire software development process. Since most software development practices apply to almost all programming languages, we can find similar essential tools for each language: code editors provide useful facilities such as syntax highlighting, code completion, code linting, and refactoring. Interactive debuggers simplify the interaction when debugging an application.

However, it is often necessary to rebuild these tools from scratch for a new programming language or to write extensions for existing IDEs in order to adapt tools. Consequently, there are often differences between tools of different languages. Sometimes they just look or feel different, sometimes they have different features or behave differently. This incidental variety has led to an incidental cognitive complexity: whenever software developers learn a new language, they also often need to learn how to use the essential tools for that language.

This problem is exacerbated in a business context where multiple developers work in a team, or even multiple development teams work on a bigger project. Then, having to learn a new language while also adapting to a new set of tools is a substantial investment for the company.

The discrepancy between different tool sets for different languages states another problem: some programming environments allow developers to do more than others. In most languages, for instance, the conventional approach to write and maintain program source code is to use offline editing tools. Afterwards, the program can then be executed. Whenever a developer wants to change the application, however, it is necessary to restart or reload the program. By contrast, other programming systems such as Smalltalk provide incremental compilation and access to language internals which encourages developers to make changes while a program is running [42, p. 22].

**The Need for Software Libraries and Frameworks**    It is good practice to design software in a modular way, because it supports extensibility as well as reusability in software systems [45, pp. 39–64]. This has led to the development of software libraries and frameworks in different languages which can be reused for different purposes.

Some languages, for example, do not have built-in support for common file formats such as the JavaScript Object Notation (JSON). Thus, developers have to implement JSON parser libraries in those languages. As another example, developers can often choose from a broad list of different server frameworks per programming language when building a server application.

However, just like tools are different for each language, frameworks and libraries for similar purposes are different. They often provide different Application Programming Interfaces (APIs), differ in features, or behave differently. Additionally, these modular software artifacts also need to be implemented from scratch for each language.

To work around this problem, many programming languages provide Foreign Function Interfaces (FFIs) which allow an application to call routines from a program written in another language. However, these interfaces are usually limited in functionality and only support to call out to the operating system, to C, or other lower-level languages. It is often inconvenient or sometimes even impossible to test the parts of an application that use FFI. But more importantly, FFI calls are commonly unsupported boundaries for tools which limits the programming experience.

**Contributions**    This work aims to solve two problems: on the one hand, we propose an approach that allows reuse of existing tools for different programming languages, rather than having to reimplement them from scratch. We claim that this makes it easier for software developers to work with more languages, as they do not have to learn how to use the different tools in each case. Instead, they get a consistent programming experience across all supported languages. On the other hand, we demonstrate how this approach also allows reuse of components written in different languages in a more convenient way. This gives developers a broader choice in terms of the frameworks and libraries they can use in their applications. Our key contributions of this paper are as follows:

- An architecture to compose multiple languages within the same live programming environment with reflective capabilities for full execution control from within the runtime.
- An approach to adapt the Smalltalk debugger and other tools, so that they work consistently across Smalltalk, Python, and Ruby.
- A convenient alternative to FFIs that allows developers to reuse software libraries and frameworks from different languages without breaking their tools.
- An experience report with different use cases to illustrate feasability, value, and limitations of our approach.

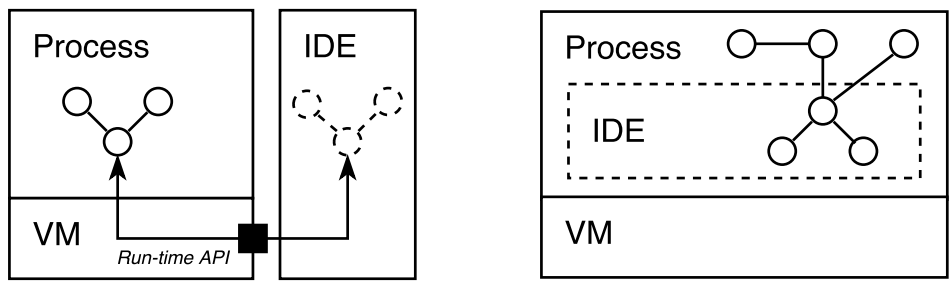

**(a)** IDEs like Eclipse or PyCharms run separately from the language process.

**(b)** In Smalltalk, the IDE is part of the process executing the language.

■ **Figure 1** Architectural comparison of two different IDE approaches.

**Outline** In Section 2, we propose and explain our approach which aims to solve different software development-related problems that were discussed before. Then, we demonstrate how we have applied this approach as part of our prototype system Squimera in Section 3. In Section 4, we give different examples of how our prototype system Squimera can be used and evaluate our approach based on these examples. Afterwards, we compare Squimera to related systems and discuss other related work in Section 5. Finally, in Section 6, we give a short summary of our approach as well as an overview of future work.

## 2 Approach

In this section, we propose an approach which aims to provide solutions for the problems described in Section 1. For this, we provide means to reuse tools as well as software libraries and frameworks.

Because there are many different types of programming languages that follow various programming paradigms, we focus on high-level dynamic programming languages for practical reasons.

### 2.1 Reuse of Existing Software Development Tools

Instead of having to build new tools, we suggest to reuse existing tools on language level. For one, this relativizes the problem for developers of having to adapt to new tools whenever they switch to another programming language. Ideally, developers are already familiar with the tools and only need to learn the syntax and concepts of another language. Additionally, this eliminates the need to write tools from scratch. Moreover, in a perfect scenario, a developer improves one tool and all other developers can benefit from that improvement, no matter which programming language they use. This perfect scenario might come with new challenges or might be hard to reach, but reusing tools that already exist is one step towards this goal.

Fabio Niephaus, Tim Felgentreff, Tobias Pape, Robert Hirschfeld, and Marcel Taeumel

**Tools Are Often Not Part of the Language**  In most programming languages, the tools are not part of the language implementation. Some languages ship tools as part of their standard library, but as long as these tools are not deeply integrated with the execution environment, they are limited in use. Some language implementers are less interested in tools or certain features, so they leave the tool building activity to language users or third parties.

The common approach of building IDEs, as shown in Figure 1a, is to run them separately from the process that executes the language. In this setup, an IDE communicates through some runtime API with the execution environment, which again often provides very different capabilities for inspecting and controlling programs that are being executed. To interact with objects at run-time, IDEs often operate on some kind of proxy objects that represent the real objects living in the language process. Many IDEs allow adaptation of their tools, but not all of their features always match those provided by the language of interest. And this mismatch often causes a rather inconvenient experience for developers.

In Smalltalk systems, however, the programming environment is part of the language and therefore runs in the same process, as illustrated in Figure 1b. Moreover, the language is mostly self-contained and designed in a generic way. Instead of having to call out to a runtime API which might have restrictions, Smalltalk tools have direct access to language internals. This allows for powerful features that support developers in building software more interactively.

We believe a Smalltalk environment makes a good reusable IDE for other programming languages. There is a renewed push in current research to allow developers to use live run-time data that can be explored and manipulated to understand and extend software systems [12, 65]. Smalltalk tools are designed for exploratory and live programming which helps to bring these styles of programming to other languages. Also, its tools are self-sustaining and mature, yet implemented concisely. Hence, they can be adapted with low effort compared with other IDEs such as Eclipse.

**Composing Language Implementations**  In order to be able to reuse tools from a Smalltalk environment, multiple languages need to be combined in one virtual execution environment. Moreover, language implementation frameworks such as Oracle's Truffle framework [69] or the RPython framework [1] are increasingly used to implement alternative execution environments for programming languages. Therefore, we can build a Virtual Machine (VM) supporting multiple languages with relatively low effort if we combine two or more language implementations written in the same framework.

However, this kind of language composition raises the question of how to execute two or more languages at the same time. Even though Smalltalk was designed many years before multi-core processors were introduced, processes are part of the language to support parallel computing [29, pp. 251–257]. Therefore, the execution of a non-Smalltalk programming language can be coordinated with the Smalltalk process scheduler when creating new Smalltalk-level processes for each foreign language invocation. Since the environment's user interface is also updated by such a Smalltalk-level process, it is possible to interact with the environment while programs written in other programming languages are being executed.

Nonetheless, the Smalltalk scheduler can only coordinate the execution of different Smalltalk-level processes if these processes complete, or, in case of a long-running process, yield at some point. Therefore, we need to ensure that the language processes can be suspended and continued at a later point in time. Since they are responsible for running code in a different interpreter loop, this implies that we need to make sure that we can suspend and continue interpreter loops at any time. This mechanism is supported by some programming languages through the notion of continuations [53], but relying on them would rule out many other languages. Instead, it would be better if the language implementation framework provides a language-independent continuation mechanism. In Truffle, Java threads could be used. In Section 3, we explain how we used delimited continuations [27] in RPython for the implementation of Squimera.

In addition to this, interpreters eventually need to suspend themselves and yield back to the main Smalltalk interpreter, so that the Smalltalk scheduler is able to switch to the next process. Some interpreters already have the capability to periodically perform specific actions. Otherwise a simple bytecode counter can be used to yield back from a non-Smalltalk interpreter after a given number of bytecodes.

Composing language implementations, on the other hand, opens up the ability to retrofit features that enable direct control over the execution of a program written in a foreign language. In case of interpreted languages, it is now possible to manipulate the interpreter loop in such a way that it is possible to restart a specific frame for example. This ability is later required to enable edit-and-continue debugging.

Moreover, programming languages can use different exception handling models. In Smalltalk, exception handling is implemented in the language, which means that the virtual machine does not need any capabilities to handle language-level exceptions. The implementation follows the resumption model of exception handling which also is needed to allow edit-and-continue debugging.

In contrast, Python, for example, uses the termination approach of error handling and does not allow developers to repair the cause of an error nor does it support to retry a failing operation [64, p. 20]. Instead, exceptions are propagated to parent execution frames which is also known as *stack unrolling* or *stack unwinding* [14]. This, however, needs to be avoided to be able to provide the same debugging experience as in Smalltalk. If the debugger is opened after the stack is unrolled, it would not be possible to see the root cause of the problem, because the corresponding frame and intermediate frames have already been removed from the call stack. But since we also do not want to change the exception handling mechanism of the Python language, we have to determine whether an exception is handled or not by one of its parent frames at the time it is raised and before the stack is modified.

In order to ensure that all of these different features exist that are required for proper tooling support and to match the Smalltalk programming experience, we can introduce appropriate abstractions for foreign language implementations. Once the abstract interface is implemented for a given language, the tools will then be ready to use for that language.

Fabio Niephaus, Tim Felgentreff, Tobias Pape, Robert Hirschfeld, and Marcel Taeumel

**Bridging Between Smalltalk and Other Languages**   This interface, however, is not only part of the virtual machine implementation. It also has to reach into the Smalltalk environment. The reason for this is that the Smalltalk tools need to communicate with objects from foreign languages, so some kind of bridge between the languages is required. First, it needs to be possible to perform foreign method or function calls from Smalltalk. For this reason, we propose to map Smalltalk semantics for message passing to other object-oriented languages. We claim that this is possible, because Smalltalk is designed in a generic way and is therefore flexible enough for semantic mappings like this. Then, we need to introduce a new Smalltalk class which is used by the VM to represent objects of a foreign language in the environment. This new Smalltalk class has to inherit from Smalltalk's `Object` class. Only this way we can ensure that foreign language objects implement the Smalltalk Meta-object Protocol (MOP) which the tools use to communicate with all objects.[1] After that, we can start overriding methods that are part of this protocol, so that they return information of the foreign language object they represent. These method overrides, however, need to return Smalltalk objects, so there also needs to be some facility that can be used to convert objects between the languages.

**Adapting the Smalltalk Tools**   After integrating another language like this into the environment, it is possible to adapt the Smalltalk tools by subclassing from the original tools. This way, we can override existing methods to make the tools work with other languages while leaving their appearance and behavior untouched.

**Architecture**   As part of our approach, we propose an architecture that we have designed during the development of our Squimera system. An overview is shown in Figure 2. The system conceptually consists of two main components: a virtual machine and a Squeak/Smalltalk environment. In this example, the VM has support for Smalltalk, Python, and Ruby. For the communication between the environment and the runtime, the virtual machine provides a plugin for each non-Smalltalk language. All of these plugins are implemented consistently as they all inherit from an abstract foreign language plugin. In Squeak/Smalltalk, we introduce different classes for each of those foreign languages. The `Python` and `Ruby` classes, in this case, facilitate the communication with the VM and can, for example, create a new `PythonProcess` or `RubyProcess` respectively. These process objects represent the execution of a language and can be used by the Smalltalk scheduler to resume them, but also to retrieve information about the current state of the execution. The classes `PythonObject` and `RubyObject`, on the other hand, are used to expose foreign objects inside the environment. Lastly, we introduce subclasses of Smalltalk tools and appropriate tool support classes.

---

[1] One could design a new, language-independent MOP for the tools to use and all languages to implement [32]. However, we suspect that any such attempt would simply create another, no less biased MOP. Since our background is influenced by object-oriented, dynamic languages, and in particular Squeak/Smalltalk's style of OOP, we decided to not hide this fact and reuse its MOP.

Live Multi-language Development and Runtime Environments

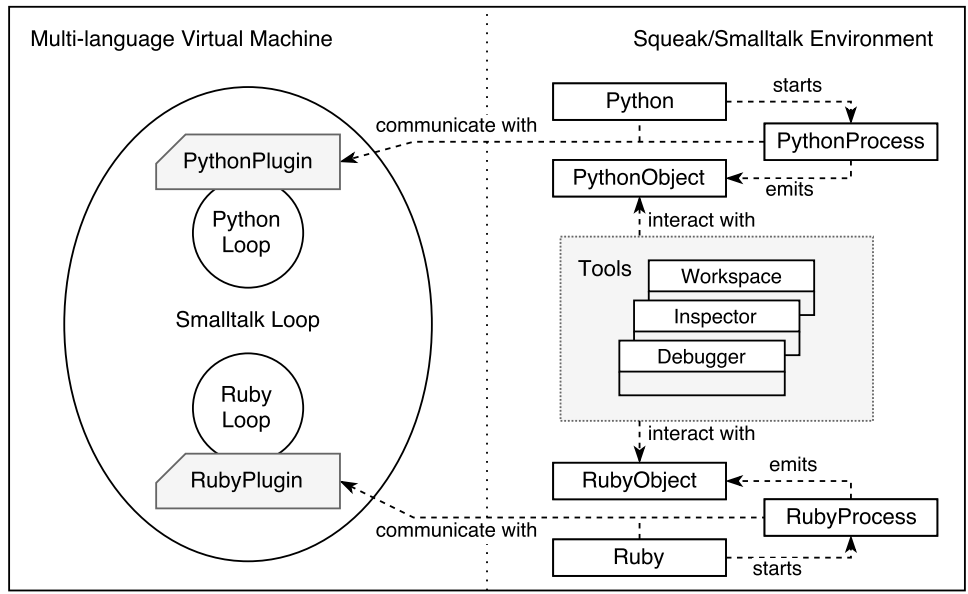

**Figure 2** Example architecture of our approach.

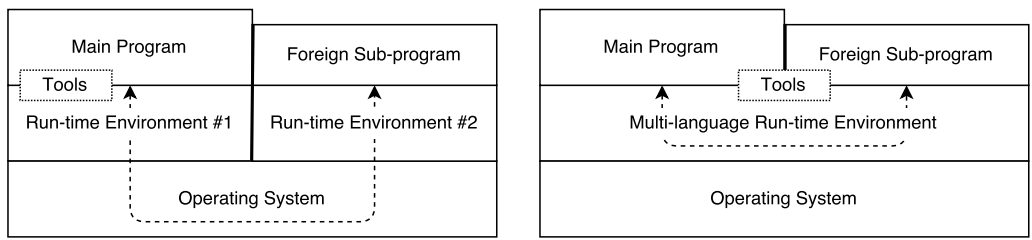

**(a)** Software reuse through FFIs.  **(b)** Software reuse through our approach.

**Figure 3** Comparing FFIs to our approach with regard to software reuse.

## 2.2 Reuse of Existing Software Libraries and Frameworks

Similar to adopting existing tools, it is valuable to developers to be able to reuse existing software artifacts. As discussed in Section 1, some languages provide FFIs which allow to call into other programs, but these interfaces are often inconvenient to use. Figure 3a demonstrates why. Usually, an FFI call is somehow propagated to the operating system which then starts a sub-program in another process on top of another execution environment. The tools, however, can only interact with the main program's runtime, which is why they are limited when it comes to debugging FFI calls for example.

On the other hand and as part of adopting the Smalltalk tools described in Section 2.1, we already had to bridge between the two languages. Therefore, it is already possible to call out to foreign languages and to convert primitive data types back and forth. But this integration of course not only allows us to adapt tools. Instead of calling internal functions of the foreign language, we can also call out to libraries that are implemented in this language. Hence, we are immediately able to reuse software libraries of a foreign language in Smalltalk. In Section 4.3, we will give a few examples of how this can be used. More importantly, the tools are able to operate on both, the main program as well as the sub-program, because everything shares the same virtual execution environment, as visualized in Figure 3b.

However, additional work is needed to support the reuse of software frameworks. Unlike in libraries, control flow is usually inverted in frameworks [24]. This implies in our case that it also needs to be possible to call from foreign languages back into Smalltalk to be able to use foreign frameworks.

Furthermore, this additional work has another positive effect. When integrating more than one foreign language with Smalltalk, for example Python and Ruby, we also are able to reuse Ruby libraries and frameworks in Python and vice-versa. The Smalltalk language then acts as a communication layer between the two languages while providing development tools for both at the same time.

## 3 Implementation

In this section, we explain how we have implemented Squimera based on the approach described in Section 2, following the architecture shown in Figure 2. The source code of Squimera is part of the RSqueak/VM and hosted on GitHub.[2]

First, we describe how we have built a multi-language virtual machine and then how we adapted Squeak/Smalltalk's tools, so that they also support other languages.

**Building a Multi-language Virtual Machine**   To build Squimera's VM, we composed different language implementations, introduced RSqueak/VM plugins that can be used to interact with other languages from Smalltalk, and lastly retrofitted runtime capabilities.

First, Squimera's VM is based on interpreter composition in RPython [1, 4]. For Python, we used PyPy [55], for Ruby Topaz [25] and the RSqueak/VM [8, 26] for Squeak/Smalltalk. Each of these language implementations consist of an interpreter loop. Since we use the tools of a Squeak/Smalltalk environment, Smalltalk is acting as the hosting language. Therefore, the VM's entry point is the interpreter loop of the RSqueak/VM.

Second, we implemented an RSqueak/VM plugin for each language to be able to execute Ruby and Python code. Each plugin can be used to evaluate code, retrieve and restart stack frames, to send messages to foreign objects, and to convert primitive objects to Smalltalk. The Smalltalk language can communicate with these plugins to request and inspect the execution of non-Smalltalk code and to forward message sends. In these cases, the VM creates and switches to a new Smalltalk-level process which holds a new execution context with a reference to a global namespace. Whenever the VM switches to such a foreign language process, the Smalltalk interpreter loop stops and the interpreter of the target language starts to execute corresponding bytecodes. After a certain number of bytecodes executed, however, the Ruby or Python interpreter yield back to the Smalltalk interpreter. This way, the Smalltalk scheduler is able to schedule another process, such as the UI process or a different language process. These processes can also be resumed because each of them has their own execution context. Similar to normal Smalltalk processes, the result of the execution is returned in case Ruby or Python code has terminated.

---

[2] https://github.com/hpi-swa/RSqueak

Third, PyPy and Topaz are both *stackful* interpreters [6, p. 390] as they manage the execution of code in nested stack frames [59]. In order to allow interpreters to yield and be resumed at any point in time, we used RPython *stacklets* [62] which are C-level coroutines [40, pp. 193–200] that can be used as a one-shot continuation. Moreover, we retrofitted Topaz and PyPy with the abilities to patch and restart frames and to detect unhandled exceptions to avoid stack unrolling. For the former, the VM creates a new frame with the corresponding changes and executes them instead of the original frame. For the latter, the VM performs bytecode analysis to check if an exception handler exists in the current call stack whenever an exception is raised. In Section 4.4, we discuss limitations of this approach. All modifications are part of the corresponding RSqueak/VM plugin. And since they implement the same interface, as described in our approach, we can ensure they provide the same runtime capabilities.

As a result, we have implemented a virtual machine with support for multiple languages. It can run a Squeak/Smalltalk programming environment which in turn can call VM internal plugins to execute Python and Ruby code. The implementation of the VM consists of less than 1,600 SLOC of additional RPython code. Of these, roughly 500 SLOC are needed for each specific language integration, while more than 500 SLOC are shared between all integrated languages.

**Adapting Squeak/Smalltalk's Tools for Non-Smalltalk Languages**  After building the VM, we continued our work in a recent Squeak/Smalltalk environment. At this point, we could have started to adapt the system browser or a text editor, so that they can provide common development features such as code completion or syntax highlighting. Instead, we decided to adapt Squeak/Smalltalk's workspace, inspector, and debugger because we wanted to focus on Squeak/Smalltalk-typical features that are missing in Ruby and Python.

To be able to execute non-Smalltalk code, we first added a class for each language, Python and Ruby. These classes implement Smalltalk's meta-object protocol and map it to Python or Ruby. As an example, `RubyObject>>#instVarNamed:`, which is used to retrieve the value of an instance variable of a Ruby object for a given name, returns `self instance_variable_get: aName`. This is equivalent to calling `self.instance_variable_get()` with a string or symbol in Ruby and demonstrates how message sends are forwarded from Smalltalk to Ruby.

After this, we adapted the Smalltalk workspace by introducing a new subclass. In this subclass, we only had to override few methods, so that the corresponding language plugin is used to evaluate foreign code. Similar to the workspace, the inspector and object explorer tools were adapted.

For syntax highlighting, we implemented a new styler which is used by all tool adaptations. This styler runs Python code on top of Squimera. It uses the Python library Pygments [13] to convert code to styled HTML which is then converted to styled

Text objects in Smalltalk. Although syntax highlighting is not as responsive as with the original Smalltalk styler, the Python-based styler works well enough and is a first example of what Squimera's multi-language runtime and programming environment allows developers to do.

Lastly, we adapted the debugger in the same way as the other tools and modified it, so that it uses the language plugins to control the execution of non-Smalltalk programs. Similar to the plugin interface on VM-level, we generalized all tool adaptations and introduced an interface for the tools, so that other languages can be added in the future.

As a result of the second part of our implementation, we have created a Squeak/Smalltalk environment with tools that support the interaction not only with Smalltalk but also with Python and Ruby code. As soon as another foreign language implements both, the VM and the in-image interfaces, the environment can provide consistent tooling for that language. The tool adaptations required around 1,250 SLOC of additional Smalltalk code, less than half of which were needed to implement language specifics, such as the meta-object protocol mapping, for both Ruby and Python.

## 4  Experience Report

In this section, we demonstrate how Squimera can be used and how it provides a consistent programming experience across all supported languages in the style of an experience report [41]. For this, we have prepared different use cases which allow us to compare the interaction with the tools when working with different languages. First, we look at Squimera's tool adaptations for live object exploration. Then, different debugging scenarios are demonstrated. Afterwards, we give an example of how libraries from different languages can be reused in Squimera. Lastly, we discuss limitations of the system and our approach.

### 4.1 Live Object Exploration

A Smalltalk-80 environment offers various tools for live data and object exploration. The Smalltalk workspace works similar to a code editor, and also allows interactive code evaluation. The Smalltalk inspector can be used to examine the internals of an object.

Figure 4a shows a Squeak/Smalltalk workspace which is used to interactively evaluate an application-specific code snippet. First, a new `DataStack` object is instantiated and stored in a variable `ds`. Then, different elements are pushed onto the stack. After that, the developer performs a *printIt* on `ds` to retrieve the object's string representation which shows the result of evaluating all previous expressions.

Next, an *inspectIt* is performed on `ds` which opens the inspector window in Figure 4b. It displays the object's string representation, but also lists an instance variable `linkedList` which the developer may inspect further. Nonetheless, this already revealed that objects of the `DataStack` class use a linked list to manage the elements internally.

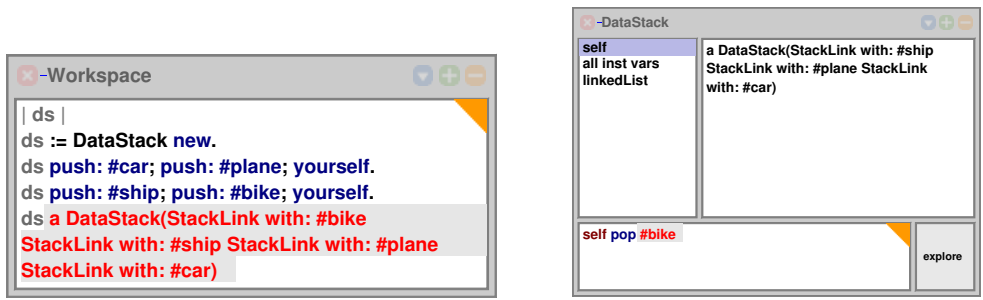

**(a)** Iteratively trying out a `DataStack` class in a Smalltalk workspace.

**(b)** After calling the `pop` method in a Smalltalk inspector.

■ **Figure 4** Inspecting an application-specific object in Squeak/Smalltalk.

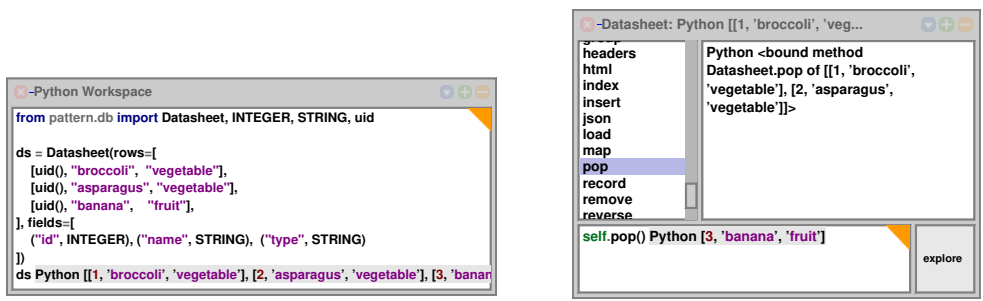

**(a)** Iteratively trying out an example of the Pattern module in a Python workspace.

**(b)** After calling the `pop()` method in a Python inspector.

■ **Figure 5** Inspecting an application-specific Python object in Squimera.

Furthermore, the `pop` method was sent to the inspected object in the Workspace-like part of the inspector. As a result, the last entry is popped off the stack and displayed as part of a *printIt*. Moreover, the string representation has also been updated to reflect the change of the object. This is a simple example to show how the inspector can provide live, immediate feedback on Smalltalk objects.

In comparison to that, Figure 5a shows Squimera's workspace in Python mode. This workspace instance is used to evaluate an example of a web mining module called Pattern [19]. The first line performs Python imports that are needed for the example. Then, a domain-specific `Datasheet` object is instantiated with different `rows` and `fields` and also stored in a variable called `ds`. Finally, the developer again performs a Smalltalk-style *printIt* on `ds` which causes the workspace to display the string representation of the Python object.

As we can see, the interaction with the Python workspace is identical to the original Squeak/Smalltalk workspace: code can be written and modified, as well as evaluated using not only *printIt*s, but also *doIt*s, *inspectIt*s, and *exploreIt*s.

When the developer inspects the object stored in `ds`, Squimera's inspector is opened on the Python object as shown in Figure 5b. The window title contains the type of the object as well as its text representation. All attributes of the object are listed similarly to how instance variables are listed in Smalltalk. This includes all special method names used by Python internally, such as `__class__` or `__dict__`, but also its public API.

This way, developers can immediately see which attributes are defined on the object. In this case, the developer has clicked on the pop attribute of the object which is a bound method.

In Figure 5b, the developer has also performed a *printIt* which executed a Python expression this time. The behavior of the tools, however, is the same: The result of the expression is displayed and the string representation of the object is updated automatically. This shows that the inspector tool also works and behaves on Python objects in the same way it behaves on Smalltalk objects.

The user experience with regard to the tools remains the same when switching to Ruby. Then, the developer can write and evaluate Ruby code in the workspace and inspect Ruby objects in exactly the same way. Figure 6 demonstrates this with the example of objects of the different languages representing the same number. The syntax to use is determined by the inspected object. Syntax highlighting adjusts automatically.

## 4.2 Debugging Experience

There are many different ways to debug a program. Interactive and live debuggers offer more functionality and flexibility compared to other debugging options, such as commandline-based post-mortem debuggers. In addition to that, the different debugging scenarios described in the following are also great examples of how well language integrations in Squimera work, because debuggers usually have to heavily interact with the underlying execution environment.

**Debugging Unhandled Exceptions**   Squimera is able to detect unhandled exceptions in Ruby and Python applications and opens a debugger accordingly. Figure 7 shows a debugger window presented to the user when an unhandled ZeroDivisionError occurs in a Python program. The window title displays the type of the Python exception as well as the error message associated with it. Moreover, there is a list of stack frames that have led to the exception. The top two frames are Python frames which is indicated by the Python icons in the list.

In this example, the exception was thrown during the execution of a Smalltalk *printIt* which is why there is no file for the code. Instead, <string> is used which is a Python convention for dynamically compiled code. Additionally, this is the reason

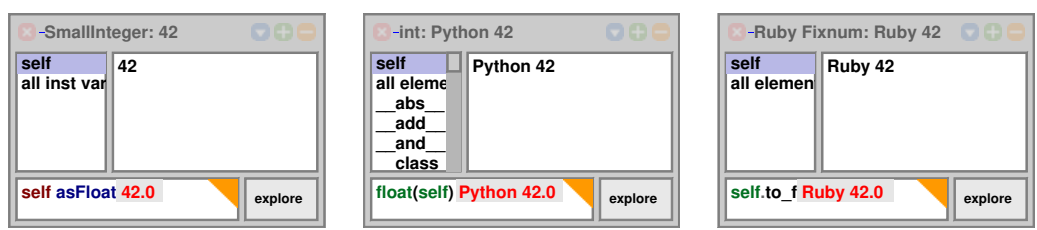

**(a)** Interaction with a Smalltalk SmallInteger object.
**(b)** Interaction with a Python int object.
**(c)** Interaction with a Ruby Fixnum object.

■ **Figure 6** Syntax and syntax highlighting are determined by the inspected object.

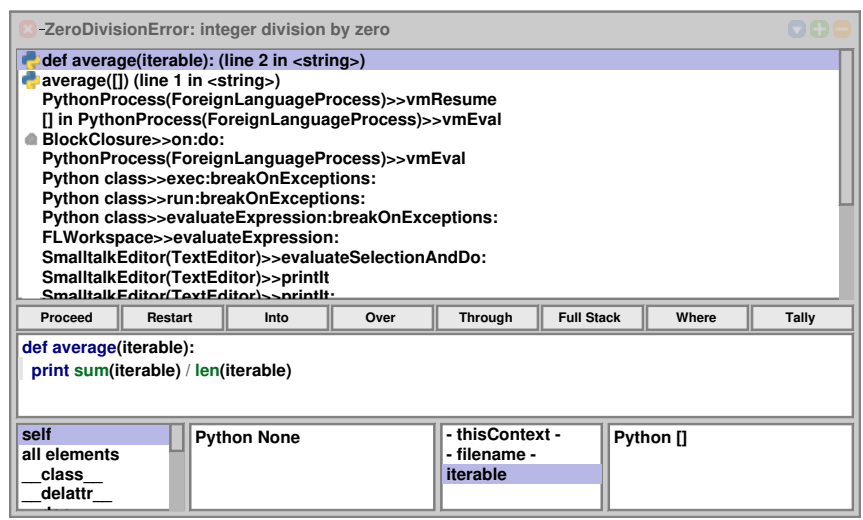

**Figure 7** Debugging a ZeroDivisionError in Python.

why there are Smalltalk frames on the call stack that, for example, perform the printIt method on a SmalltalkEditor. The code editor widget of the debugger shows the code being executed in the selected frame with syntax highlighting enabled. The current line of execution is also highlighted, just like in an original Smalltalk debugger. Since the code has been executed in the global Python context, the left half of the bottom lists the Python globals. The right half provides more information on the Python context attached to the frame: - *thisContext* - is a reference to the frame object and - *filename* - displays the current filename. Then, the local variables are listed. Since there is only the variable iterable, the developer is immediately able to identify the root cause of the problem. The naively implemented average function does not check if the iterable is empty, and in this case a ZeroDivisonError is thrown.

At this point, the developer may further inspect stack frames in order to examine why an empty iterable was provided. It is also possible to modify code and then restart a specific frame. Or the developer may decide to carry on with the execution by pressing the *Proceed* button. An unhandled exception would then be returned as the result of the execution.

**Interrupting Running Applications** Another way of debugging applications in Smalltalk is related to its live exploration capabilities. At any point in time, a developer may trigger a user interrupt [28, p. 409] to open a debugger on the currently running process. This technique is especially useful to understand the internal state of a long-running program, such as a server application or an emulator. Since Python and Ruby programs are being executed as part of a Smalltalk-level process, it is possible to interrupt them in Squimera, just like any other Smalltalk process.

For demonstration purposes, we run Optcarrot, a NES emulator written in Ruby [22], with Squimera. After a couple of seconds, the emulator opens a new window and starts to draw a NES ROM. At this stage, we press the interrupt key which triggers a user interrupt. Since the Ruby process is taking up the most computing resources

Fabio Niephaus, Tim Felgentreff, Tobias Pape, Robert Hirschfeld, and Marcel Taeumel

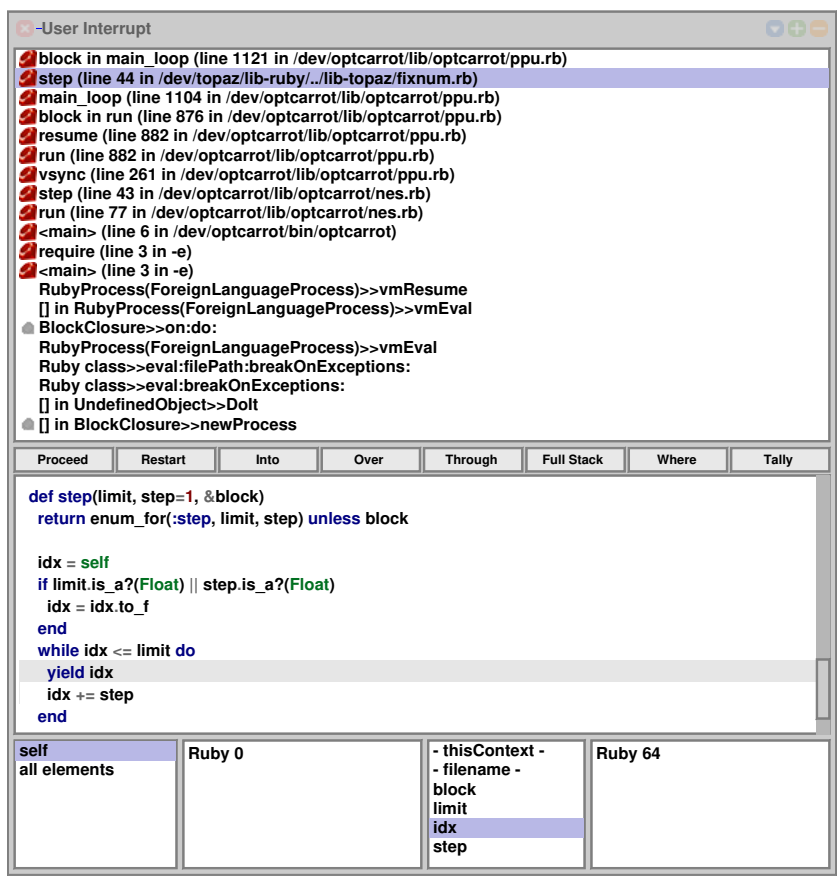

**Figure 8** Debugging a Ruby process while it is running.

at this moment, it is likely that the user interrupt happens during the execution of the emulator. Otherwise, we could also interrupt it using Squeak/Smalltalk's process browser.

Then, a debugger windows, as shown in Figure 8, is opened and the Ruby process is no longer scheduled for execution. Therefore, the execution of the NES emulator is suspended.

Ruby stack frames are highlighted with a Ruby icon, similar to how Python frames were highlighted in Figure 7. This time, there are many Ruby stack frames, most of them, including the top frame, execute code from a file called ppu.rb, in which the Picture Processing Unit of the NES is implemented. Therefore, we can observe that the emulator was refreshing its display at the time the interrupt was triggered.

In the example, the second frame is selected which executes code from the Fixnum class which is part of Ruby's standard library. The Ruby code from line 1104 in the ppu.rb file, that caused a new frame in fixnum.rb, is o.step(248, 8) do. Therefore, the value of self in the execution context is 0, while limit is 248 and step is 8. The value of idx at that point is 64. All this information is accessible with a few clicks in the debugger. Following the stack frames even further, we find the Ruby <main> frame. This frame is followed by the first Smalltalk frame which is responsible for resuming the language process. The last but one frame contains the *doIt* that initially started the execution of

**Live Multi-language Development and Runtime Environments**

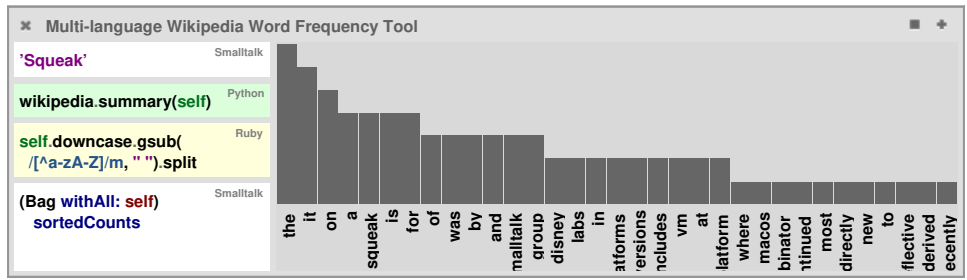

**Figure 9** A word frequency tool built with Vivide, written in Smalltalk, Python, and Ruby.

Optcarrot followed by the bottom frame which created a new Smalltalk-level process for this.

Hence, the debugger can be used to inspect and modify the running emulator including frames executing its own code, but also code from the standard library, as well as Squimera's code that is used to execute the Ruby language process and also Smalltalk code that initiated everything.

## 4.3 Reusing Software Libraries

As discussed in Section 1, developers build modular software which is crucial to support extensibility and reusability. However, the reusability aspect is often limited by language boundaries, because developers can only reuse software written in their program's language. In Section 2.2, we explained how our approach allows not only reuse of software development tools, but also of software libraries and frameworks in a convenient way. We illustrate this with the following example.

Figure 9 shows a tool to measure the word frequency of Wikipedia summaries. It is built in the Vivide framework [60] and consists of different Workspace-like boxes. Each box can be configured to evaluate one of the supported languages by Squimera and is connected with the next box passing its result along, similar to IPython notebooks [51]. The search term is entered in the first Smalltalk box. Then, this is passed as a Python string bound to the keyword self into the next box which calls the summary function of a Python library Wikipedia [30]. The result is again passed to the next box, this time as a Ruby string. In Ruby, a list of lowercase words is then extracted. Lastly, this list is passed back to Smalltalk where it is put into a bag to count word frequencies before it is visualized in a bar chart on the right.

The Vivide framework in Smalltalk allows us to compose UI elements and connect them to build this example tool. Instead of having to understand Wikipedia's API, we can use a Python library which wraps around it and provides various convenience methods. We leverage the regular expression engine of Ruby to extract a list of lowercase words from the summary. And finally, we use Squeak/Smalltalk's Bag API to count and sort these words. Therefore, this example demonstrates how Squimera allows developers to choose libraries and frameworks from different languages to build applications. Furthermore, Squimera is able to provide a consistent debugging experience across all these languages in case of an exception thrown in any part of this example tool.

### 4.4 Limitations of Squimera

There are some limitations with regard to our approach and to our prototype system Squimera. Due to time constraints, we were unable to fully implement the ability to call Smalltalk methods from foreign languages. As discussed in Section 2.2, this would allow us to reuse software frameworks, which is currently not supported in Squimera. Additionally, it is possible that a non-Smalltalk interpreter blocks in some cases, for example when a web server is using blocking sockets. Then, the programming environment freezes because the execution cannot switch back to the Smalltalk interpreter updating the UI until the blocking interpreter yields.

Altough performance was not a main focus when building Squimera, a ballpark measurement of the Richards benchmark [54] in our Python integration suggests that the performance of Python programs can be similar to CPython in Squimera, even though we use the PyPy interpreter. Therefore, one could say that Smalltalk-level processes may decrease the performance almost as much as performance can be increased by PyPy's Just-in-time (JIT) compiler. On the other hand, a full-fledged IDE is now running at the same time and in the same operating system process. However, this preliminary result should be taken with a grain of salt, as performance highly depends on the number of active Smalltalk processes as well as the number of bytecodes a non-Smalltalk interpreter is allowed to process at a time. At the moment, the default number of bytecodes is set to $10,000$, but it is possible to adjust this value when starting the VM. A higher number increases performance of the corresponding interpreter while it can cause the development environment to become less responsive depending on the running program.

As part of Section 2.1, we explained that Squimera needs to be able to detect unhandled exceptions to avoid stack unwinding in languages that use the termination model of exception handling. The model is comparably easy to implement and yet powerful during the execution of a language. Nonetheless, it is rather hard to reliably detect unhandled exceptions and therefore to provide useful debugging facilities. The detection in Squimera performs bytecode analysis. Other IDEs, such as PyCharm [36], perform their analysis on source code level which has no connection to actual execution state. Both approaches, however, are error prone and cannot cover all cases. One example are the many different ways to define try-except statements or to mask exceptions with builtins in Python. Python is too dynamic in that sense. The same holds true for Ruby.

**Conceptual Mismatches of Languages and Tools**   In addition to these shortcomings, there are further limitations of our approach with regard to the languages and tools in general that we now discuss in more detail.

First of all, we have only integrated object-oriented programming languages as part of Squimera. These languages are relatively similar to Smalltalk, which makes it, for instance, easier to map semantics and to conceptually reuse Smalltalk tools. Nonetheless, we do not believe that it is impractical to integrate languages following other programming paradigms. But this requires additional work.

However, we noticed that there are also some conceptual problems with regard to the Smalltalk tools with our integration of Python and Ruby. These may also occur when integrating other languages. One example is code management. Since everything is an object in Smalltalk, code is represented by objects. Source code for Python, Ruby, and many other languages, on the other hand, is managed in files on the file system. Therefore, adapting tools like the system browser which are used to implement and maintain an application also needs additional work.

According to Cunningham [68], one lesson learned from VisualAge for Java is that only people with Smalltalk experience enjoy the tools. One important reason for that is that the tools are written in Smalltalk, so Smalltalk experience is required to change them. The ability to call Smalltalk from other languages might yet again help with this problem. When this is possible, we could implement a wrapper library for Squeak/Smalltalk's `ToolBuilder` API, so tools can be built and modified in other languages and without having to write any Smalltalk code.

Furthermore, the ability to call Smalltalk from foreign languages also has a downside. Currently, Ruby and Python methods can be called with Smalltalk objects as arguments. The VM currently converts primitive data types automatically to corresponding objects of the target language, otherwise it will pass on the original object. Therefore, once the other way is also possible, it is necessary to also provide converting methods for each integrated language, for example `asRuby` or `asPython`. When using a foreign framework, for example, it needs to be possible to create objects in the framework's language that can be passed back to it accordingly. Additionally, it is possible to turn off the automatic conversion. Then, developers have more control about the objects that are being passed around. On the other hand, they need to think more about the origins of different objects and manually perform type conversions when necessary. This, in turn, is additional work and could also result in less concise code.

## 5  Related Work

In this chapter, we discuss solutions and technologies related to our work. Further, we compare them with our approach and with the Squimera system.

### 5.1  Tools and Integrated Development Environments

**Integrated Development Environments for Python and Ruby**   Compared to our prototype system Squimera, IDEs for Python and Ruby, such as PyCharm [36], the Wing Python IDE [70], and RubyMine [37], as well as IDEs with support for multiple-languages, such as Eclipse [56] and Visual Studio [48], provide a lot more and mature features for Ruby and Python development, as they have been in active development by larger communities and companies for years. As explained in Section 2, however, all of these IDEs are based on the architecture illustrated in Figure 1a. Even though most of them support different Ruby and Python interpreters, their tools are limited to the capabilities of the used runtime.

On the other hand, the Squimera programming environment runs in the same process that also executes the languages, following the Smalltalk architecture as shown in Figure 1b. This gives more control over the execution of different languages, as it supports to restart frames and to patch code at run-time for example. This in turn enables features such as edit-and-continue debugging in a consistent way in Squimera.

**IBM VisualAge**   IBM VisualAge was a family of mainly Smalltalk-based IDEs with support for different programming languages. In VisualAge for Java, a Smalltalk VM was used with support for both, Smalltalk and Java [16].

This demonstrated that it is possible to use a Smalltalk environment for tools as well as a multi-language VM for the execution of another language in a commercial context. Instead of being a family of IDEs, Squimera is a single IDE with support for multiple languages and with consistent tooling.

**Smalltalk/X**   STX:LIBJAVA [35] and SmallRuby [66] are experimental Java and Ruby implementations on top of Smalltalk/X. The former project uses a VM with support for Java bytecode instructions, while the latter compiles Ruby code to Smalltalk/X bytecode. In addition, both projects allow interoperability with Smalltalk.

Similar to Squimera, SmallRuby and STX:LIBJAVA provide Smalltalk-based development tools for Ruby and Java. Objects of these two languages expose the same meta-object protocol as Smalltalk objects which ensures that the original tools can operate on them. This way, the Smalltalk/X debugger, for example, also supports mixed stacks. Squimera also aligns the meta-object protocol for foreign objects, but runs non-Smalltalk languages in Smalltalk-level processes.

**Language Server Protocol and Monto**   Microsoft's Language Server Protocol (LSP) [47] allows language support to be implemented independently of any IDE. For this, each languages has to provide a persistent *language server* that any code editor or IDE supporting the protocol can then communicate with over network requests. This provides common features such as code completion, go-to-definition, or syntax highlighting. Keidel, Pfeiffer, and Erdweg have presented Monto [38] which follows a similar goal, however, their protocol allows language servers to be stateless and to work without having to maintain a copy of the source code.

In Squimera, the IDE is deeply integrated with the runtime, which has an interface that is implemented by each language. While the protocols aim to be IDE agnostic, Squimera focuses on a single tool set. Currently, Squimera's language features are limited to syntax highlighting, object inspection tooling, and edit-and-continue debugging. The latter two are not yet supported by the LSP, although it is being worked on [49].

**Multi-language Debugger Architecture**   Vran and Píše proposed an architecture that combines different debuggers into one debugger with multi-language support [67]. For this, they suggest a generic debugging interface as an abstraction layer which is then used by a multi-language debugger.

Live Multi-language Development and Runtime Environments

As part of our approach, we presented an interface on the level of the execution environment instead. This not only allows developers to use only one debugger in a similar way, but also ensures that the different interpreters support the same capabilities. This in turn is a requirement to be able to provide a consistent debugging experience with support for edit-and-continue debugging. In contrast, their architecture only allows the intersection of all debugging features provided for all languages.

**Blink**   Lee, Hirzel, Grimm, and McKinley have presented Blink, a debugger composition that works across Java, C, and the Jeannie programming language [43]. In order to debug multiple languages, it uses an agent that interposes on language transitions, so that it can control and reuse existing language-specific debuggers.

The debugger provided by Squimera, on the other hand, is an adapted version of a Smalltalk debugger that can debug through different languages by communicating with a consistent VM interface.

**Eco**   Diekmann and Tratt presented an approach for syntax-directed-style editors [20]. Their prototype editor Eco allows developers to write composed programs and supports Python, HTML, and SQL at AST level. The editor uses *language boxes* for code written in the different languages. While Eco ensures that the language composition will parse correctly, its runtime integration and tooling is minimal compared with Squimera.

### 5.2  Cross-language Integration Techniques

**Foreign Function Interfaces and Inter-process Communication**   With foreign function interfaces, it is possible to call out to programs written in other languages. They are used by developers not only to reuse existing software, but also to speed up performance-critical computations by calling out to more efficient and often lower-level code. Java, for example, supports FFI calls through the Java Native Interface [44], while Squeak/Smalltalk, Python and many other languages base their FFI capabilities on libffi [31].

Inter-process communication can be use for similar purposes, but allows applications to communicate with each other. Microservices, for example, use a network connection and communicate through Remote Procedure Calls (RPCs) [50].

Squimera provides a new mechanism for software reuse which we believe is superior in terms of usability to approaches based on FFIs and RPCs. Since all subroutines run in the same operating system process, objects from different languages can be directly accesses and tools, such as the debugger, have more control over the execution.

**Language Interoperability**   In the last decades, a lot of work and research has been done to provide means that allow native interaction between multiple languages.

In 1998, for example, Cleary, Kohn, Smith, and Smolinski presented an idea which enables language interoperability for high-performance scientific applications through an Interface Definition Language [17].

Similar to that, Hamilton described how the Common Language Runtime (CLR) enables the integration of programming languages on runtime level [33]. For this, all compatible languages including C#, Java, C++, and Python have to be compiled to the Common Intermediate Language (CIL) which can then be executed on by the CLR. Further, Eaddy and Feiner have presented an edit-and-continue implementation that allows runtime updates to programs running on CLR [21]. However, the edit-and-continue feature is currently only supported for C++, C#, and Visual Basic [46], although it is said to also work for IronPython according to the paper.

Moreover, Evans and Verburg have shown how the JVM allows for interoperability between Java and other languages that run on top of the JVM such as Scala or Clojure [23].

The CLR approach requires language implementations that can compile a language into an intermediate language. For language interoperability with the JVM, it needs to be possible to interpret all languages on top of it. Squimera, on the other hand, makes use of different language implementations instead and runs a different interpreter for each language.

**Interpreter Composition in RPython**   With Unipycation [4], Barrett, Bolz, and Tratt presented a composition of interpreters for Python and Prolog in RPython, based on PyPy and Pyrolog [10]. Although their bi-language VM is built similarly to Squimera's VM, it does not have any further scheduling mechanisms for switching between interpreters. Also, their work mainly focuses on performance advantages of a JIT compiler rather than on the programming experience or on tooling.

Moreover, Barrett, Bolz, and Tratt later compared Unipycation with other approaches to interpreter composition [3]. As a result, they concluded that their approach not only led to a well-performing VM, but also was comparatively easy to implement.

For similar reasons, we used interpreters written in RPython to build Squimera. Instead of having to write language implementations from scratch or compose interpreters in C, we were able to reuse existing language implementations with relatively low effort.

**Truffle's Ployglot Engine**   The Truffle framework in combination with the GraalVM provides high-performance language interoperability capabilities [32, 71]. The different language implementations, including TruffleJS, TruffleC, and TruffleRuby, emit Abstract Syntax Tree (AST) nodes that Truffle can execute and optimize. Furthermore, it is possible to mix AST nodes from different languages at run-time through an API provided as part of its Polyglot engine.

Although some work has been done to provide debugging support for languages implemented in Truffle [58, 63], the framework focuses on performance rather than on tooling. In contrast, our approach can be seen as a tool-first approach, as our focus is on providing a better programming experience. Nonetheless, we believe it is possible to build or adapt tools that work with different Truffle languages similar to Squimera. It is, however, unclear to what extent the framework is able to provide a consistent programming experience as this would also require the same advanced runtime capabilities, such as hot code patching, to be available across all Truffle languages.

### 5.3 Live Programming

**Terminology**   The term *live* is used diversely in the field of programming and usually refers to techniques that create the impression of changing a program while it is running. In *live coding*, for example, a program is changed while it is running as part of a performance of audio or visual art [7, 18]. The liveness is necessary to change the output in-time.

Techniques for *live programming* are designed to help developers to understand the behavior of their code at development time [34]. In particular, some of these techniques aim to minimize feedback loops by, for example, continuously displaying runtime state [15].

Interactive programming environments such as Smalltalk-80 aim at evolving long-running systems [28, 57]. Therefore, they allow developers to explore, inspect, and modify the behavior and state of objects at run-time while also supporting developers with mechanisms for short feedback loops.

Since Squimera is based on Squeak/Smalltalk, it relates to the latter category of liveness. Squimera does not introduce new tools or mechanisms for live programming. Instead, it demonstrates that it is possible to make Smalltalk tools for live objects inspection, exploratory programming, and online debugging work with other languages and without having to reimplement any of the live mechanisms. However, some of these mechanisms, such as edit-and-continue debugging, rely on runtime capabilities, such a hot-code swapping, that we had to retrofit as part of building Squimera's VM.

**Interactive Programming in Ruby and Python**   Ruby and Python provide Read–eval–print Loops (REPLs) that can be used to interactively run code. However, developers can only evaluate code and see the result which is comparable to Smalltalk *printIts*. Moreover, it is rather inconvenient to inspect objects, as it usually takes more effort to drill down on the right aspects of an objects in a REPL than by using tools like the Smalltalk inspector. More importantly, the Smalltalk tools can provide live feedback, so developers can observe changes without requiring further interaction.

Moreover, there are different IDE extensions, such as the live-py-plugin [39] for PyDev and Eclipse, that enable live capabilities to some extent for various dynamic languages. However, these systems usually reload programs entirely and therefore cannot provide a higher level of liveness than Tanimoto's level 3 [61]. On the other hand, Squimera allows the modification of Ruby and Python programs at run-time and is able to provide visual feedback in a few hundreds of milliseconds as it is based on Squeak/Smalltalk [52].

## 6   Conclusion

In this paper, we have described two common problems with regard to reuse developers encounter when working on software: language-specific tooling and other language-bound software artifacts. First, they often need to use specific tools for the language they develop in. This requires them to acquire special knowledge which, in some cases, is language-specific and cannot be applied when developing in other languages. This problem scales in a business context, where several developers work in teams on a complex software product. In this case, switching to a different language not only requires to migrate code to the new language. More importantly, developers need to be trained to be able to work efficiently with the language's development tools. This can be a substantial financial investment for companies. Seconds, similar software libraries and frameworks often have to be recreated in each programming language. A reason for this is that concepts, ideas, or even entire architectures that have proven to work well in one language may also work well in others.

In this paper, we presented an approach which attempts to address both of these problems through reuse by leveraging a multi-language runtime environment. Reusing tools for software development not only reduces the work for language or tool implementers. More importantly, it also makes it easier for developers, as they can work with familiar tools to develop software in different languages in a uniform way. As a result of that, software libraries can also be reused in a more convenient way compared to traditional FFI-based approaches. Our prototype system Squimera demonstrates that it is possible to reuse live programming tools of a Smalltalk environment for other high-level dynamic languages and to create a more consistent programming experience.

**Future Work** There are multiple avenues for future work. As discussed in Section 4.4, it would be useful if foreign languages could call to Smalltalk. This would not only allow the proper reuse of software frameworks in Smalltalk, but also to reuse Python libraries and frameworks in Ruby and vice-versa.

Currently, we have mainly focused on runtime tools in Squimera. But we want to adapt more Smalltalk tools, such as the system browser or the test runner which can be used to write and maintain program code.

Further, we want to find out if we can integrate more languages into Squimera, especially languages based on programming paradigms other than object-orientation. There are many RPython-based implementations of various languages that can be used, for example for Prolog [4], Racket [5], Lisp [2], or SQL [9].

Lastly, we want to investigate if our approach can also be applied to other language ecosystems. The GraalVM, for instance, also supports language interoperability of languages implemented in Truffle. Also, it would be interesting to see if it is possible to adapt tools from Eclipse or the NetBeans IDE [11], so that they behave more consistently across different languages.

**Acknowledgements** We would like to thank Carl Friedrich Bolz-Tereick, Armin Rigo, and the PyPy team for their help with the PyPy interpreter. We gratefully acknowledge the financial support of HPI's Research School[3] and the Hasso Plattner Design Thinking Research Program.[4]

---

[3] https://hpi.de/en/research/research-school.html
[4] https://hpi.de/en/dtrp/

**Live Multi-language Development and Runtime Environments**

**About the authors**

**Fabio Niephaus** is a PhD student in the Software Architecture Group of the Hasso Plattner Institute at the University of Potsdam. His research interests include programming languages, virtual execution environments, and software development tools. Contact Fabio at fabio.niephaus@hpi.uni-potsdam.de.

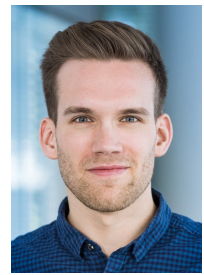

**Tim Felgentreff** is a Senior Software Engineer at Oracle Labs and a former member of the Software Architecture Group of the Hasso Plattner Institute at the University of Potsdam, where he received his PhD. His research interests include programming-language extensions and high-performance dynamic-language virtual machines. Contact Tim at tim.felgentreff@hpi.uni-potsdam.de.

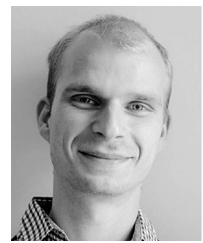

**Tobias Pape** is a PhD student in the Software Architecture Group of the Hasso Plattner Institute at the University of Potsdam. He is interested in virtual-machine construction, language design, and data structure optimization. Contact Tobias at tobias.pape@hpi.uni-potsdam.de.

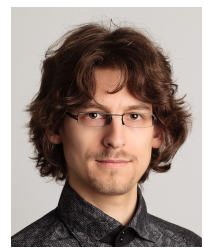

**Robert Hirschfeld** leads the Software Architecture Group of the Hasso Plattner Institute at the University of Potsdam. His research interests include dynamic programming languages, development tools, and runtime environments to make live, exploratory programming more approachable. Hirschfeld received a PhD in computer science from Technische Universität Ilmenau. Contact Robert at robert.hirschfeld@hpi.uni-potsdam.de.

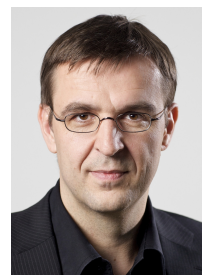

**Marcel Taeumel** is a PhD student in the Software Architecture Group of the Hasso Plattner Institute at the University of Potsdam. His research activities include the area of building graphical tools for programmers, especially data-driven ones for program comprehension tasks. Contact Marcel at marcel.taeumel@hpi.uni-potsdam.de.

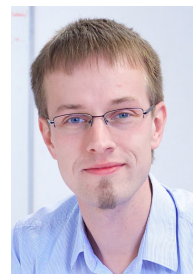